\documentclass[twoside]{article}
\usepackage{fleqn,espcrc2,epsf}
\usepackage{graphicx}
\usepackage[figuresright]{rotating}

\newcommand{\blambda}{{\overline{\lambda}}}
\newcommand{\nn}{\nonumber}
\newcommand{\be}{\begin{equation}}
\newcommand{\ee}{\end{equation}}
\newcommand{\bea}{\begin{eqnarray}}
\newcommand{\eea}{\end{eqnarray}}
\newcommand{\tr}{{\rm Tr}}
\newcommand{\hmu}{{\hat{\mu}}}

\title{SUSY Ward identities in $N$=1 SYM theory on the lattice}

\author{Federico Farchioni\address[DESY]{Deutsches Elektronen-Synchrotron, DESY,
       \\
        Notkestr. 85, D-22603 Hamburg, Germany}\thanks{Talk given by Federico Farchioni.}, 
        Alessandra Feo\address[UM]{Institut f\"ur theoretische Physik,
        Universit\"at M\"unster,
       \\
        Wilhelm-Klemm-Str. 9, D-48149 M\"unster, Germany},
        Tobias Galla\address[UO]{Department of Physics,
        University of Oxford, Theoretical Physics, 
       \\
        1 Keble Road, Oxford OX1 3NP, UK},
        Claus Gebert\addressmark[DESY], Robert Kirchner\addressmark[DESY],
       \\
        Istv\'an Montvay\addressmark[DESY], Gernot M\"unster\addressmark[UM],
        Anastassios Vladikas\address[URTV]{INFN, Sezione di Roma 2 c/o Dipartimento di Fisica,
        Universit\'a di Roma ``Tor Vergata'',
       \\
        Via della Ricerca Scientifica 1, I-00133 Rome, Italy},
       \\
        [0.5em]
        DESY-M\"unster-Roma Collaboration \\[0.5em] }
       
\begin{document}

\begin{abstract}

 The SUSY Ward identities (WIs) for the $N$=1 $SU(2)$ SUSY
 Yang Mills theory discretized on the lattice with Wilson 
 fermions (gluinos) are considered. The study is performed in 
 the framework of a Monte Carlo simulation of the model 
 with light dynamical gluinos. The renormalization and mixing constants of the lattice 
 SUSY current $Z_S$ and $Z_T$ and the additively renormalized gluino mass $m_S$
 are unknown parameters of the SUSY WIs. Using suitable on-shell 
 combinations of the WIs, the ratios $Z_T/Z_S$ and $m_S/Z_S$ 
 are determined non-perturbatively at one value of the coupling constant $g_0$ 
 and two values of the hopping parameter $\kappa$.
\vspace{1pc}
\end{abstract}

\maketitle

\section{INTRODUCTION}

The non-perturbative regime of SUSY gauge theories is
highly interesting, since among other things it may provide
a possible mechanism for dynamical SUSY-breaking.
An increased understanding of the non-perturbative phenomena of SUSY
gauge theories can be achieved in the framework of the lattice regularization.
An immediate difficulty arises however because the lattice regularized
theory is not supersymmetric, since the Poincar\'e invariance (which is a
sector of the superalgebra) is lost.

The present computational resources allow now to approach the 
simplest of the SUSY gauge theories, the $N=1$ supersymmetric Yang Mills (SYM) models, 
where the $N_c^2$-1 gluons are accompanied by neutral fermionic partners 
(gluinos) in the same adjoint representation of the color group.
The Wilson discretization of the SYM models, proposed by
Curci and Veneziano \cite{CuVe}, explicitly breaks SUSY by the Wilson term 
and, softly, by a gluino Majorana mass term.
SUSY is only recovered upon tuning the hopping parameter to the massless 
gluino limit and by taking the continuum limit. 

An extensive analysis of the low-energy behavior of the $N$=1 $SU(2)$ SYM model 
with Wilson fermions was performed by this collaboration 
in a Monte Carlo simulation with light dynamical gluinos, see \cite{KietAl,CaetAl} 
and references therein. The issue of the restoration of SUSY in the massless 
gluino limit is now addressed, as proposed in \cite{DoetAl},
by the non-perturbative analysis of the on-shell lattice  SUSY WIs. 
The same issue, from a perturbative point of view, has been dealt with 
in another contribution \cite{FaetAl}.

An approach to the SYM  models with domain wall fermions 
was recently considered in \cite{Col}.

\section{LATTICE FORMULATION}

In the Curci-Veneziano approach \cite{CuVe}, which we adopt,
the standard Wilson discretization is applied to the SYM model,
where SUSY and the (anomalous) $U(1)$ chiral symmetry
of the continuum theory are explicitly broken by 
the Wilson term and a gluino Majorana mass term;
SUSY is also broken by the space-time lattice. 
The issue of the recovering of the appropriate symmetry 
in the continuum limit is addressed through the discussion of the relevant WIs.

The gluonic action is the standard plaquette action
\be  \label{gauge}
S_g  = \beta\, \sum_{pl}                            
(1 - \frac{1}{N_c} {\rm Re\,Tr\,} U_{pl}) \ ,   
\ee
where the bare gauge coupling is given by $\beta \equiv 2N_c/g_0^2$.
The fermionic action is (with Wilson parameter $r$=1)
\begin{eqnarray}  \label{fermi}
S_f&\!\!\!\!=\!\!\!\!&
\sum_x \tr \bigl[
 \frac{1}{2}\blambda_x(-1+\gamma_\mu)
 U^\dagger_{x\mu}\lambda_{x+\hmu} U_{x\mu}
\nn
\\
&\!\!\!\!+\!\!\!\!&\frac{1}{2}\blambda_{x+\hmu} (-1-\gamma_\mu)
U_{x\mu}\lambda_x U^\dagger_{x\mu} 
\nn\\
&\!\!\!\!+\!\!\!\!&(m_0+4)\blambda_x\lambda_x\bigr]\ ,
\end{eqnarray}
$\lambda\equiv\sum_r T_r\lambda^r$ 
being a Majorana field (gluino field) transforming according 
to the adjoint representation of the gauge group
($T_r$ are the generators).

\subsection{SUSY WARD IDENTITIES}

The effect of the explicit breaking
of the chiral symmetry by the Wilson fermions is well understood in 
the framework of lattice QCD \cite{BoetAl}, where 
the WIs of the continuum can be recovered through
renormalization of the chiral currents and the quark masses.
The lattice SUSY WIs of the SYM models are derived in \cite{CuVe} 
using the same theoretical setup.
Specifically, the  SUSY current $S_{\mu}(x)$, considered in on-shell correlation
functions, undergoes multiplicative renormalization and mixing with a gauge-invariant 
current $T_{\mu}(x)$; the renormalized current reads
\be\label{renormcur}
\hat{S}_{\mu}(x)\:=\:Z_S\:S_{\mu}(x)\:+\:Z_T\:T_{\mu}(x)\ ,
\ee
which, for a gauge invariant local operator ${\cal O} (y)$ ($y \neq x$), 
satisfies the integrated WIs 
\bea\label{renormward}
\nn
\sum_{\vec{x}}\langle\nabla_0 \hat{S}_{0}(x)\,{\cal O}(y)\rangle&\!\!\!\!=\!\!\!\!&m_S\! 
\sum_{\vec{x}}\langle\chi(x) {\cal O}(y)\rangle + O(a)   \\
&&
\eea
(the exact expressions for $S_\mu(x)$, $T_\mu(x)$, $\chi(x)$ are given below).
The quantity $m_S$ is obtained from $m_0$ by additive renormalization,
$m_S\equiv m_0-\bar{m}_S(m_0,g_0)$. 
As it is evident from the WIs (\ref{renormcur},\ref{renormward}), 
the renormalizations $Z_S$, $Z_T$ and $\bar{m}_S$
reabsorb the main lattice artifact of the explicit SUSY breaking; 
the residual $O(a)$ breaking terms on the r.h.s. of 
(\ref{renormward}) imply indeed effects vanishing exponentially with $g_0$.
If these are negligible, the condition $m_S$=0 ensures
the restoration of SUSY on the lattice.
When $m_S$=0, also the renormalized gluino mass
multiplying the soft breaking term in the lattice chiral WIs 
vanishes \cite{CuVe} as expected.

For the numerical analysis of the SUSY WIs 
we consider two possible lattice forms of the SUSY and mixing currents;
a point-like discretization:
\bea\label{dis1}
\nn
S^{(1)}_{\mu}(x) &\!\!=\!\!&
-\sum_{\rho\sigma} \sigma_{\rho\sigma} \gamma_\mu
 \tr \bigl[
 P^{(cl)}_{x,\rho\sigma} \lambda_x
\bigr]
\\
T^{(1)}_{\mu}(x) &\!\!=\!\!&
2\sum_{\nu} \gamma_\nu
 \tr \bigl[
 P^{(cl)}_{x,\mu\nu} \lambda_x
\bigr]\ ,
\eea
and a point-split one \cite{Tani}:
\bea\label{dis2}
\nn
S^{(2)}_{\mu}(x) &\!\!=\!\!&
-\frac{1}{2} \sum_{\rho\sigma} \sigma_{\rho\sigma} \gamma_\mu
 \tr \bigl[
 P^{(cl)}_{x,\rho\sigma} U^\dagger_{x\mu}\lambda_{x+\hmu}U_{x\mu}
\\
\nn
&\!\!+\!\!&P^{(cl)}_{x+\mu,\rho\sigma} U_{x\mu}\lambda_x U^\dagger_{x\mu}
\bigr]
\\
\nn
T^{(2)}_{\mu}(x) &\!\!=\!\!&
\sum_\nu \gamma_\nu 
 \tr \bigl[
 P^{(cl)}_{x,\mu\nu} U^\dagger_{x\mu}\lambda_{x+\hmu}U_{x\mu}
\\
&\!\!+\!\!&P^{(cl)}_{x+\mu,\mu\nu} U_{x\mu}\lambda_x U^\dagger_{x\mu}
\bigr]\ .
\eea
The values of $Z_S$ and $Z_T$ are expected to be different
for these two choices. 
Consistently with parity and time-reversal,
$P^{(cl)}_{x,\mu\nu}$ is defined as a 
clover-symmetrized lattice field tensor; 
moreover, the lattice derivative 
in (\ref{renormward}) is defined, for the discretization (\ref{dis1}): 
$\nabla^{(1)}_\mu f(x)\equiv 1/2(f(x+\hmu)-f(x-\hmu))$; 
in the case (\ref{dis2}): $\nabla^{(2)}_\mu f(x)\equiv f(x)-f(x-\hmu)$.

Finally, the lattice form of the operator $\chi(x)$ in
the soft-breaking  term is
\be\label{chi}
\chi(x) = \sum_{\rho\sigma} \sigma_{\rho\sigma}
 \tr \left[P^{(cl)}_{x,\rho\sigma} \lambda_x\right]\ .
\ee

\section{NUMERICAL ANALYSIS}

\begin{figure}[t]
\vspace*{-5pt}
\epsfxsize=6cm
\epsfbox{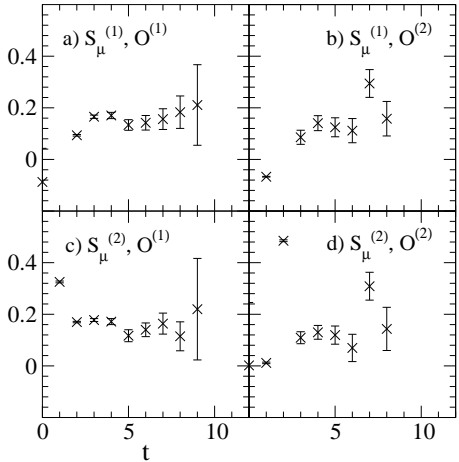}
\vspace*{-25pt}
\caption{$m_S/Z_S$ as a function of the time-separation $t$ at $\kappa$=0.1925.}
\label{fig:mu}
\vspace*{-10pt}
\end{figure}

\begin{figure}[t]
\vspace*{-5pt}
\epsfxsize=6cm
\epsfbox{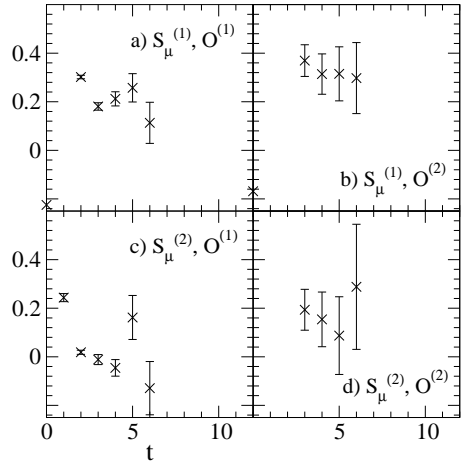}
\vspace*{-25pt}
\caption{$Z_T/Z_S$ as a function of the time-separation $t$ at $\kappa$=0.1925.}
\label{fig:zeta}
\vspace*{-10pt}
\end{figure}

We use two sets of gauge configurations at $\beta$=2.3 on a $12^3\times24$ lattice,
at $\kappa$=0.1925 and at $\kappa$=0.194 with a statistics
of 4212 and 2034 configurations, respectively.
The second set is closer to the critical point of the chiral phase transition
$\kappa_c$=0.1955(5) as determined in \cite{KietAl}.
All configurations were obtained by applying the two-step dynamical-fermion
multi-bosonic algorithm, see \cite{CaetAl} and references therein for details. 

We consider for our study the WIs obtained by inserting in (\ref{renormward})
the two gauge invariant operators
\bea\label{O1}
{\cal O}^{(1)}(x)&=&\sum_{i<j}\sigma_{ij}\tr[P_{x,ij}\lambda_x]
\\
\label{O2}
{\cal O}^{(2)}(x)&=&\sum_{i}\sigma_{0i}\tr[P_{x,0i}\lambda_x]\ ,
\eea
with the clover and the simple-plaquette definition of the lattice field tensor $P_{x,\mu\nu}$
(the latter only for the operator (\ref{O1})). The above operators  
undergo a (common) multiplicative renormalization and the form (\ref{renormward}) 
of the WIs is consequently not spoiled. 
Since spatial integration breaks hypercubic symmetry in (\ref{renormward})
these two operator insertions amount to independent choices.

Because of the Dirac structure of the operators involved, 
Eq.~(\ref{renormward}) gives 16 (real) WIs 
for every time-separation $t\equiv x_0-y_0$
and a given insertion operator ${\cal O}(x)$.
Due to the Majorana nature of the field $\lambda_x$
and the discrete symmetries of the action (\ref{gauge},\ref{fermi}),
only two equations are however independent.
Consequently, determinations of the ratios $m_S/Z_S$ and $Z_T/Z_S$ can be obtained,
for a fixed ${\cal O}(x)$ and for every $t$, by solving a two-by-two linear system.
These determinations contain $O(a)$ effects.

In order to get a significant signal for the correlations, 
a combined APE smearing for the gluon field and Jacobi 
smearing for the gluino field was performed on the insertion operator.
We use two sets of smearing parameters (set A and B) giving equally good results
for the ratios.

In Figs. \ref{fig:mu} and \ref{fig:zeta} 
the determinations of the ratios $m_S/Z_S$ and $Z_T/Z_S$ 
are reported, at $\kappa$=0.1925 as a function of the time-separation $t$,
for the two discretizations of the SUSY and mixing currents
(\ref{dis1}) (\ref{dis2}) and the two insertion operators (\ref{O1}) (\ref{O2})
(with clover field tensor).
A plateau can be observed for $m_S/Z_S$ (Fig.~\ref{fig:mu}) 
for $t\geq 3$ with the insertion operator ${\cal O}^{(1)}(x)$ and
for $t\geq 4$ with ${\cal O}^{(2)}(x)$: in these regions of $t$ 
contact terms due to the time-extension of the operators entering the correlations
are absent. The signal for 
$Z_T/Z_S$ (Fig.~\ref{fig:zeta}) is noisier.

\begin{table*}[htb]
\caption{Results for  $m_S/Z_S$ and $Z_T/Z_S$ with the discretizations 1 and 2
of the currents and the two insertion operators, with plaquette and clover
field tensor and smearing parameters A and B.}
\label{tab:res}
\newcommand{\m}{\hphantom{$-$}}
\renewcommand{\tabcolsep}{.4pc} 
\begin{tabular}{@{}lcccllll}
\hline
\multicolumn{1}{c}{$\kappa$} & current & field tensor & smearing  &
\multicolumn{1}{c}{$m_S/Z_S$ (${\cal O}^{(1)}$)} & 
\multicolumn{1}{c}{$m_S/Z_S$ (${\cal O}^{(2)}$)} & 
\multicolumn{1}{c}{$Z_T/Z_S$ (${\cal O}^{(1)}$)} & 
\multicolumn{1}{c}{$Z_T/Z_S$ (${\cal O}^{(2)}$)} \\ 
\hline
0.1925 & 1 & plaquette    &   A   & 0.155(11)      &           &\m 0.184(28)   &             \\
0.1925 & 1 & plaquette    &   B   & 0.173(11)      &           &\m 0.160(25)   &             \\
0.1925 & 1 & clover       &   A   & 0.170(13)      & 0.144(18) &\m 0.244(35)   &\m 0.29(6)   \\
0.194  & 1 & clover       &   A   & 0.118(17)      &            &\m 0.190(45)     &             \\
\hline
0.1925 & 2 & plaquette    &   A   & 0.152(9)       &           &  $-$0.010(33)   &    \\
0.1925 & 2 & plaquette    &   B   & 0.168(9)       &           &  $-$0.044(32)   &             \\
0.1925 & 2 & clover       &   A   & 0.170(12)      & 0.132(16) &  $-$0.018(43)   &\m 0.11(7)   \\
0.194  & 2 & clover       &   A   & 0.130(15)      &            &  $-$0.03(5)     &         \\
\hline
\end{tabular}
\end{table*}

The results of the global fit over the range of time-separations $t\geq4$
are reported in Table~\ref{tab:res}.

Different  discretizations of the SUSY or mixing current 
give different renormalizations $Z_S$ and $Z_T$ for any finite $g_0$. 
In particular, the discretization (\ref{dis2})
gives a mixing coefficient $Z_T$ compatible
with zero according to our data, but in case of
(\ref{dis1}) $Z_T$ seems to be small and non-zero.

Results from the two independent insertion operators 
${\cal O}^{(1)}(x)$ and ${\cal O}^{(2)}(x)$ differ by $O(a)$ effects.
The discrepancy appears to be comparable to the statistical uncertainty,
even if the insertion operator ${\cal O}^{(2)}(x)$ seems to give a lower
value for $m_S/Z_S$ compared to ${\cal O}^{(1)}(x)$.
Observe that also results coming from different lattice forms 
of a given operator differ by discretization errors.

Theoretically, $m_S/Z_S$ should decrease
when the hopping parameter approaches the expected critical point $\kappa_c$, 
and vanish at $\kappa_c$; $Z_T/Z_S$ should take its asymptotic
value for massless gluino. Comparison between data at $\kappa=0.1925$ and
$\kappa=0.194$ seems to confirm this expectation.
The statistical indetermination still prevents
an accurate extrapolation of $\kappa$ to massless gluino.

\section{CONCLUSIONS}

The present study shows that the extraction of the ratios
$m_S/Z_S$ and $Z_T/Z_S$ from the on-shell SUSY Ward identities
is technically feasible with the computing resources at hand.
The main technical difficulty (related to SUSY) is that 
high-dimensional operators with a mixed gluonic-fermionic composition 
must be considered, introducing relatively large statistical fluctuations.
This difficulty can be handled with an appropriate smearing
procedure. The non-perturbative determination of the
ratio $m_S/Z_S$ can be used for 
an independent extrapolation to the critical hopping parameter 
corresponding to massless gluinos.
Our results for $m_S/Z_S$ and $Z_T/Z_S$ are consistent 
with the WIs (\ref{renormcur},\ref{renormward}) with $O(a)$
effects comparable to the statistical indetermination.

\vspace*{.5em}

 {\bf Acknowledgement:}
 The numerical study presented here has been performed on the
 CRAY-T3E computers at John von Neumann Institute for Computing (NIC),
 J\"ulich.
 We thank NIC and the staff at ZAM for their kind support.
 We also thank Giancarlo Rossi for stimulating discussions.

\end{document}